\documentclass[aps, twocolumn, showpacs, amsmath]{revtex4-2}
\usepackage{dcolumn}
\usepackage{bm}
\usepackage{graphicx}
\usepackage{color}
\usepackage{amsthm}
\usepackage{amssymb}
\usepackage{hyperref}

\begin{document}
\def\be{\begin{equation}}
\def\ee{\end{equation}}

\def\bc{\begin{center}}
\def\ec{\end{center}}
\def\bea{\begin{equation}}
\def\eea{\end{equation}}
\newcommand{\avg}[1]{\langle{#1}\rangle}
\newcommand{\Avg}[1]{\left\langle{#1}\right\rangle}

\def\ie{\textit{i.e.}}
\def\etal{\textit{et al.}}
\def\m{\vec{m}}
\def\G{\mathcal{G}}

\newcommand{\davide}[1]{{\bf\color{blue}#1}}
\newcommand{\gin}[1]{{\bf\color{green}#1}}

\title{Random walks on complex networks with first-passage resetting}

\author{Feng Huang{$^{1,3}$}}

\author{Hanshuang Chen{$^2$}}\email{chenhshf@ahu.edu.cn}

\affiliation{$^1$Key Laboratory of Advanced Electronic Materials and Devices \& School of Mathematics and Physics, Anhui Jianzhu University, Hefei, 230601, China \\ $^2$School of Physics and Materials Sciences, Anhui
University, Hefei, 230601, China \\$^3$Key Laboratory of Architectural Acoustic Environment of Anhui Higher Education Institutes, Hefei, 230601, China}

\begin{abstract}
We study discrete-time random walks on arbitrary networks with first-passage resetting processes. To the end, a set of nodes are chosen as observable nodes, and the walker is reset instantaneously to a given resetting node whenever it hits either of observable nodes. We derive exact expressions of the stationary occupation probability, the average number of resets in the long time, and the mean first-passage time between arbitrary two non-observable nodes. We show that all the quantities can be expressed in terms of the fundamental matrix $\textbf{Z}=(\textbf{I}-\textbf{Q})^{-1}$, where $\textbf{I}$ is the identity matrix and $\textbf{Q}$ is the transition matrix between non-observable nodes. Finally, we use ring networks, 2d square lattices, barbell networks, and Cayley trees to demonstrate the advantage of first-passage resetting in global search on such networks.  
\end{abstract}

\maketitle

\section{Introduction}

Random walks on complex networks are not only the core of studying stochastic dynamical process on networked systems \cite{masuda2017random,klafter2011first}, but also pave a broad range of applications, such as community detection \cite{rosvall2008maps,zhou2004network,pons2005computing}, epidemic spreading \cite{RevModPhys.87.925,colizza2007reaction,PhysRevX.1.011001}, human mobility \cite{PhysRevE.86.066116,riascos2017emergence,barbosa2018human}, ranking and searching on the web \cite{PhysRevLett.92.118701,newman2005measure,lu2016vital,kleinberg2006complex,RevModPhys.87.1261}. In this context, two of important quantities can be identified. One is the occupation probability at stationarity, which quantifies the frequency of visiting each node in the long time \cite{PhysRevLett.92.118701,PhysRevE.87.012112}. The other one is the mean first-passage time (MFPT) from one node to another, which is closely relevant to many important applications \cite{redner2001guide}, such as epidemic extinction \cite{WKBReview1,PhysRevLett.117.028302}, neuronal firing \cite{tuckwell1988introduction}, consensus formation \cite{PhysRevLett.94.178701}, and so on \cite{bray2013persistence}. For the standard random walks on undirected and unweighted networks, it was established that the stationary occupation probability of a node is directly proportional to its degree, i.e., the number of its neighboring nodes \cite{PhysRevLett.92.118701}. The MFPT between two nodes can be calculated by some different methods \cite{redner2001guide,van1992stochastic}, such as the renewal method \cite{PhysRevLett.92.118701,PhysRevE.87.012112,zhang2011mean,PhysRevE.79.021127}. It has been shown that the MFPT is related to the eigenmodes of the transition matrix besides the one corresponding to the largest eigenvalue that gives the stationary information. That is to say, the MFPT is also related to relaxation properties of random walks \cite{PhysRevLett.92.118701,PhysRevE.87.012112,PhysRevLett.109.088701}.

Recently, random walks subject to resetting processes have received increasing attention \cite{evans2011diffusion,evans2011diffusion2,pal2016diffusion} (see \cite{evans2020stochastic} for a review). The walker is interrupted stochastically and then reset to a given position. The stochastic reset has lead to many intriguing results. The resetting can drive the system to a nonequilibrium steady state. An infinite MFPT can become finite, and an optimal
resetting rate was found that makes the MFPT minimal. These nontrivial findings render to an enormous recent activities in the field, both in theory \cite{evans2014diffusion,pal2017first,chechkin2018random}, experiments \cite{tal2020experimental,besga2020optimal} and applications \cite{gupta2014fluctuating,rotbart2015michaelis,fuchs2016stochastic,pal2017integral,gupta2020work,basu2019symmetric,magoni2020ising}. In particular, in a recent paper, Riascos \textit{et al.} studied random walks on arbitrary networks subject to a constant resetting rate \cite{PhysRevE.101.062147}. They derived the exact expressions of the stationary probability distribution and the MFPT. Subsequently, the results are generalized to the case of multiple resetting nodes \cite{arXiv:2104.00727}. Wald and B\"ottcher introduced a framework for studying classical, quantum, and hybrid random walks with stochastic resetting on arbitrary networks, in which they derived analytical solutions of the occupation probability for a classical or quantum random walker \cite{PhysRevE.103.012122}.

Very recently, Bruyne \textit{et al.} introduced a first-passage resetting process in one-dimensional space \cite{de2020optimization,de2021optimization}. The particle is reset to the starting position whenever it reaches a specified threshold. Contrary to standard resetting, the time at which first-passage resetting occurs is defined by the motion of the diffusing particle itself, rather than being imposed externally. They showed that in an infinite domain this process is nonstationary and its probability distribution exhibits rich features. In a finite domain, they defined a nontrivial optimization in which a cost is incurred whenever the particle is reset and a reward is obtained while the particle stays near the position to which the particle is reset. They derived the condition to optimize the net gain in this system, namely, the reward minus the cost. 

In the present work, we study discrete-time random walks on arbitrary networks with \textit{first-passage resetting} processes. To the end, we firstly choose a set of nodes as observable nodes, and a node as the only resetting node. Whenever the diffusing particle reaches either of observable nodes, it is reset instantaneously to the resetting node and the random walk process is restarted. We exactly derive the expressions of the occupation probability at stationarity, the average number of resetting as a function of time, and the MFPT between two nodes on arbitrary networks. The results show that all the three quantities can be expressed by the so-called fundamental matrix, that is the inverse of $\textbf{I}-\textbf{Q}$, in which $\textbf{I}$ is the identity matrix and $\textbf{Q}$ is the transition matrix between non-observable nodes (including the resetting node). In particular, we demonstrate our results on ring networks, two-dimensional square lattices, barbell networks, and finite-size Cayley trees. On such networks, we find that the first-passage resetting is advantageous to global search processes.

\section{Model}
\begin{figure}
	\centerline{\includegraphics*[width=0.6\columnwidth]{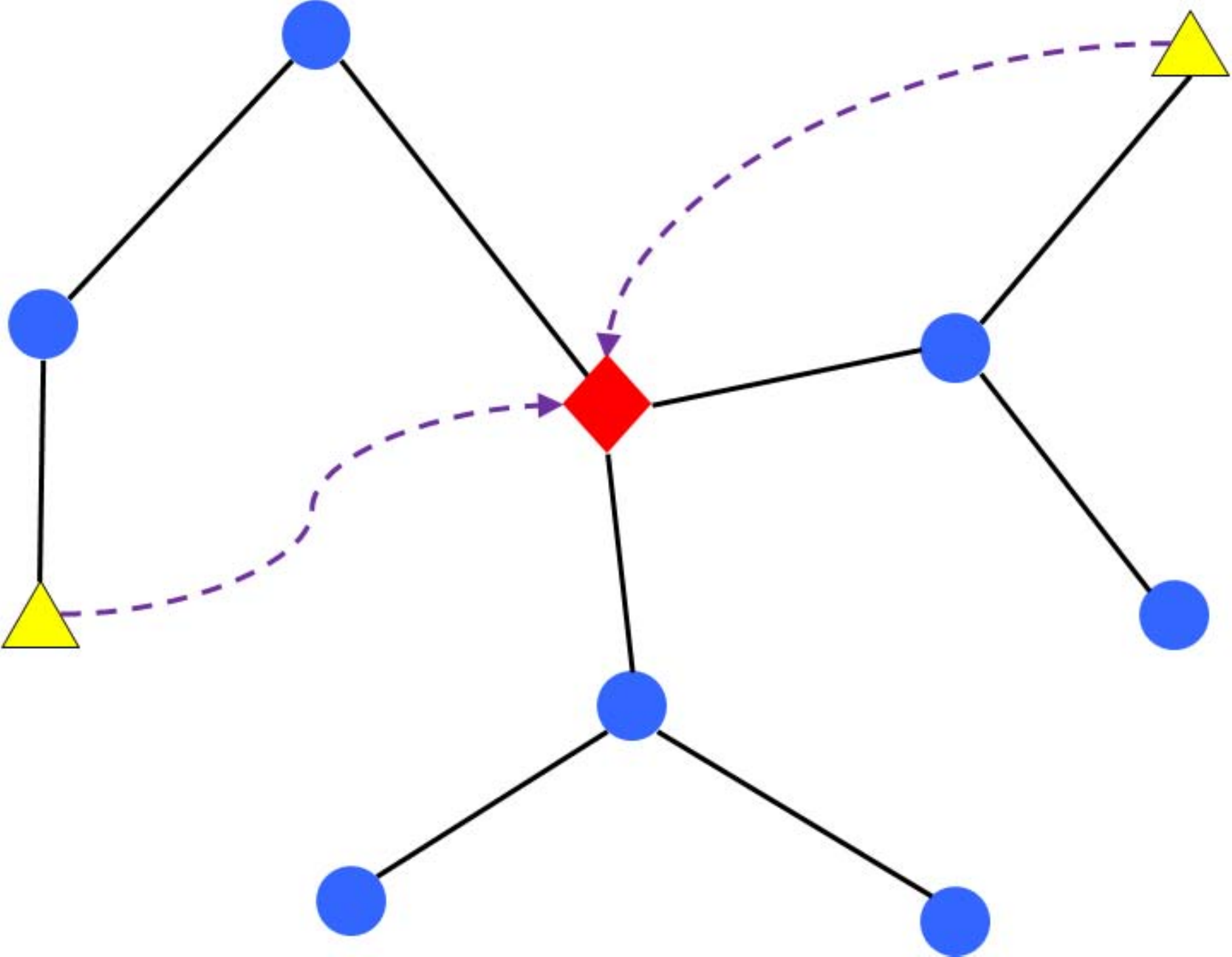}}
	\caption{An illustration of random walks on a network with first-passage resetting. The walker is reset instantaneously to the resetting node ($\diamond$) whenever it arrives at either of observable nodes ($\triangle$), and then the random walk process  is proceeded. }\label{fig1}
\end{figure}

We consider a particle that performs discrete-time random walks on a network consisted of $N$ nodes, from which we choose $m$ observable nodes, labelled with $\textbf{o} = \left\{ {{o_1}, \cdots ,{o_m}} \right\}$, and a single node as the resetting node, labelled with $r$. Assuming that the particle is located at a non-observable node $i$ at time $t$, the particle hops to one of its neighboring nodes, saying node $j$, at time $t+1$, with the probability $1/d_i$, where $d_i$ is the degree of node $i$. If node $j$ is either of observable nodes, the particle is then reset instantaneously to the resetting node $r$, see Fig.\ref{fig1} for an illustration. Note that the resetting node $r$ can not be either of observable nodes.

For convenience, let us denote by $\textbf{Q}$ the $n \times n$ transition matrix between non-observable nodes ($n=N-m$), whose entry $Q_{ij}$ denotes the transition probability from the non-observable node $i$ to the non-observable node $j$, and by $\textbf{R}$ the $n \times m$ transition matrix from non-observable nodes to observable nodes, whose entry $R_{ij}$ denotes the transition probability from the non-observable node $i$ to the observable node $o_j$.

\section{Stationary Occupation Probability}
Let $P_{ij}(t)$ be the probability that the particle starts from node $i$ at time $t=0$ and arrives at node $j$ at time $t$,  which satisfies the following master equation, 
\begin{equation}\label{eq0}
{P_{ij}}\left( {t + 1} \right) = \sum\limits_{k = 1}^n {{Q_{kj}}{P_{ik}}\left( t \right)}  + {\delta _{jr}}\sum\limits_{l = 1}^m {\sum\limits_{k = 1}^n {{R_{kl}}{P_{ik}}\left( t \right)} } .
\end{equation}
The first term on the right-hand side of Eq.\ref{eq0} represents hops between non-observable nodes whereas the second term describes the first-passage resetting from observable nodes to the node $r$ ($\delta _{jr}$ denotes the Kronecker delta).

Instead of beginning from the master equation one can write down a renewal equation,
\begin{equation}\label{eq1}
{P_{ij}}\left( t \right) = {G_{ij}}\left( t \right) + \sum\limits_{t' = 0}^t {{F_{i\textbf{o}}}\left( t' \right){P_{rj}}\left( {t - t'} \right)}.
\end{equation}
Note that both nodes $i$ and $j$ are non-observable nodes, and $F_{i\textbf{o}}(0)=0$. $G_{ij}(t)$ is the probability of starting from node $i$ and arriving at node $j$ at time $t$ without hitting any observable node up to time $t$, given by
\begin{equation}\label{eq3}
{G_{ij}}\left( t \right) = {\left( {{\textbf{Q}^t}} \right)_{ij}}.
\end{equation}
$F_{i\textbf{o}}(t)$ is the first-passage probability of starting from node $i$ and hitting any observable node at time $t$, which can be written as
\begin{equation}\label{eq2}
{F_{i\textbf{o}}}\left( t \right) = \sum\limits_{j = 1}^m {{F_{i{o_j}}}\left( t \right)},
\end{equation}
where $F_{i{o_j}}(t)$ is the first-passage probability of starting from node $i$ and hitting the observable node $o_j$ at time $t$, given by
\begin{equation}\label{eq4}
{F_{i{o_j}}} (t) = {\left( {{\textbf{Q}^{t - 1}}\textbf{R}} \right)_{ij}}.
\end{equation}
Note that $F_{i{o_j}}(0)=0$ since the starting node $i$ does not belong to the set of observable nodes.

Performing the discrete-time Laplace transform for Eq.(\ref{eq1}), $\tilde f\left( s \right) = \sum\nolimits_{t = 0}^\infty  {{e^{ - st}}f\left( t \right)}$, we have
\begin{equation}\label{eq5}
{{\tilde P}_{ij}}\left( s \right) = {{\tilde G}_{ij}}\left( s \right) + {{\tilde F}_{i\textbf{o}}}\left( s \right){{\tilde P}_{rj}}\left( s \right),
\end{equation}
where $\tilde{G}_{ij}(s)$ and $\tilde{F}_{i\textbf{o}}(s)$ can be obtained from Eq.(\ref{eq3}) and Eq.(\ref{eq4}), given by
\begin{equation}\label{eq6}
{{\tilde G}_{ij}}\left( s \right) = \sum\limits_{t = 0}^\infty  {{e^{ - st}}{{\left( {{\textbf{Q}^t}} \right)}_{ij}}}  = {\left[ {{{\left( {{\textbf{I}_n} - {e^{ - s}}\textbf{Q}} \right)}^{ - 1}}} \right]_{ij}},
\end{equation}
and
\begin{equation}\label{eq7}
{{\tilde F}_{i{o_j}}}\left( s \right) = \sum\limits_{t = 1}^\infty  {{e^{ - st}}{{\left( {{\textbf{Q}^{t - 1}}\textbf{R}} \right)}_{ij}}}  = {\left[ {{{\left( {e^{  s}} {{\textbf{I}_n} - \textbf{Q}} \right)}^{ - 1}}\textbf{R}} \right]_{ij}},
\end{equation}
where $\textbf{I}_n$ is the identity matrix of dimension $n$. 
Letting $i=r$ in Eq.(\ref{eq5}), we have
\begin{equation}\label{eq8}
{{\tilde P}_{rj}}\left( s \right) = \frac{{{{\tilde G}_{rj}}\left( s \right)}}{{1 - {{\tilde F}_{r \textbf{o}}}\left( s \right)}}.
\end{equation}
Substituting Eqs.(\ref{eq6},\ref{eq7},\ref{eq8}) into Eq.(\ref{eq5}), we obtain
\begin{widetext}
\begin{align}\label{eq9}
	{{\tilde P}_{ij}}\left( s \right) &= {{\tilde G}_{ij}}\left( s \right) + \frac{{{{\tilde F}_{i\textbf{o}}}\left( s \right)}}{{1 - {{\tilde F}_{r\textbf{o}}}\left( s \right)}}{{\tilde G}_{rj}}\left( s \right) \nonumber \\ &=  {\left[ {{{\left( {{\textbf{I}_n} - {e^{ - s}}\textbf{Q}} \right)}^{ - 1}}} \right]_{ij}}+  \frac{{\sum_{k = 1}^m {{{\left[ {{{\left( {{e^s}{\textbf{I}_n} - \textbf{Q}} \right)}^{ - 1}}\textbf{R}} \right]}_{ik}}} }}{{1 - \sum_{k = 1}^m {{{\left[ {{{\left( {{e^s}{\textbf{I}_n} - \textbf{Q}} \right)}^{ - 1}}\textbf{R}} \right]}_{rk}}} }}{\left[ {{{\left( {{\textbf{I}_n} - {e^{ - s}}\textbf{Q}} \right)}^{ - 1}}} \right]_{rj}}.
\end{align}
\end{widetext}

The stationary occupation probability can be obtained by evaluating the limit
\begin{equation}\label{eq10}
{P_{j}}\left( \infty  \right) = \mathop {\lim }\limits_{s \to 0} \left( {1 - {e^{ - s}}} \right){\tilde P_{ij}}\left( s \right).
\end{equation}
In the left-hand side of Eq.\ref{eq10}, we have omitted the subscript for the starting node $i$ since  the information on the starting node has been erased in the limit of long time.
Since ${{\tilde F}_{i\textbf{o}}}\left( 0 \right) = \sum_{t = 0}^\infty  {{F_{i\textbf{o}}}} \left( t \right) = 1$, the denominator in the second term of Eq.(\ref{eq9}) is equal to zero in the limit of $s \to 0$, and we then apply the L'H\^opital rule to calculate the limit, which leads to (see Appendix \ref{appen1} for details)
\begin{equation}\label{eq11}
{P_j}\left( \infty  \right) = \frac{{{Z_{rj}}}}{{\sum\nolimits_{k = 1}^n {{Z_{rk}}} }},
\end{equation}
where we have defined the matrix
\begin{equation}\label{eq12}
\textbf{Z} = {\left( {\textbf{I}_n - \textbf{Q}} \right)^{ - 1}}.
\end{equation}
$\textbf{Z}$ is called the \textit{fundamental} matrix whose entry $Z_{rj}$ denotes the average time spent on the node $j$ starting from the resetting node $r$ before the particle hits either of the observable nodes. The denominator in Eq.(\ref{eq11}) is the requirement of normalization, $\sum\nolimits_{j = 1}^n {{P_j}\left( \infty  \right)} =1$.

\section{Average number of resets}
An important quantity in the resetting process is the average number of resets up to time $t$, $\mathcal{N}(t)$, which satisfies a backward renewal equation,
\begin{equation}\label{eq13}
\mathcal{N}\left( t \right) = \sum\limits_{t' = 0}^t {{F_{r\textbf{o}}}\left( {t'} \right)} \left[ {1 + \mathcal{N}\left( {t - t'} \right)} \right].
\end{equation}
Eq.(\ref{eq13}) accounts for the particle first hitting either of the observable nodes at
any time $t' \leq t$, then the process is renewed over the
time range $t-t'$, so there will be on average $1+\mathcal{N}(t-t')$
resets. Taking the Laplace transform of Eq.(\ref{eq13}) then leads to
\begin{equation}\label{eq14}
\mathcal{\tilde N} \left( s \right) = \frac{{{{\tilde F}_{r\textbf{o}}}\left( s \right)}}{{\left( {1 - {e^{ - s}}} \right)\left[ {1 - {{\tilde F}_{r\textbf{o}}}\left( s \right)} \right]}}.
\end{equation}
Substituting Eq.(\ref{eq7}) into Eq.(\ref{eq14}), we obtain
\begin{equation}\label{eq15}
\mathcal{\tilde N} \left( s \right) = \frac{{\sum_{j = 1}^m {{{\left[ {{{\left( {{e^s}{\textbf{I}_n} - \textbf{Q}} \right)}^{ - 1}}\textbf{R}} \right]}_{rj}}} }}{{\left( {1 - {e^{ - s}}} \right)\left\{ {1 - \sum_{j = 1}^m {{{\left[ {{{\left( {{e^s}{\textbf{I}_n} - \textbf{Q}} \right)}^{ - 1}}\textbf{R}} \right]}_{rj}}} } \right\}}}.
\end{equation}
Taking the limit of $s \to 0$, we obtain $\mathcal{\tilde N} \left( s \right) = \sigma {\left( {1 - {e^{ - s}}} \right)^{ - 2}}$, and then the long-time behavior of the average number of resetting events (see Appendix \ref{appen2} for details), 
\begin{equation}\label{eq16}
\mathcal{N } \left( t \right) =\sigma t,
\end{equation}
where the growth rate $\sigma$ is given by
\begin{equation}\label{eq17}
\sigma =\frac{1}{{\sum_{j = 1}^n {{Z_{rj}}} }}.
\end{equation}

\section{Mean First-Passage Time}
Let $F_{ij}(t)$ be the first-passage probability of arriving at a non-observable node $j$ at time $t$ starting from a non-observable node $i$ at time $t=0$. The connection between $F_{ij}(t)$ and $P_{ij}(t)$ is expressed as \cite{PhysRevLett.92.118701}
\begin{equation}\label{eq18}
{P_{ij}}\left( t \right) = {\delta _{ij}}{\delta _{t0}} + \sum\limits_{t' = 0}^t {{F_{ij}}\left( t' \right)} {P_{jj}}\left( {t - t'} \right),
\end{equation}
where the Kronecker delta symbol insures the initial condition $P_{ij}(0)=\delta_{ij}$ ($F_{ij}(0)$ is set to zero). In the Laplace domain, Eq.(\ref{eq18}) becomes
\begin{equation}\label{eq19}
{{\tilde F}_{ij}}\left( s \right) = \frac{{{{\tilde P}_{ij}}\left( s \right) - {\delta _{ij}}}}{{{{\tilde P}_{jj}}\left( s \right)}}.
\end{equation}
Supposing that there is a trap located at node $j$, the particle is absorbed immediately once the particle arrives at the trap. Let us denote by $S_{ij}(t)$ the probability that the particle has not been absorbed by node $j$ up to time $t$, providing that the particle starts from node $i$ at time $t=0$. The connection between the survival probability $S_{ij}(t)$ and first-passage probability $F_{ij}(t)$ is given by $S_{ij}(t) = 1 - \sum\nolimits_{t' = 0}^t {F_{ij}(t')}$. From this connection, we have  $F_{ij}(t)=S_{ij}(t-1)-S_{ij}(t)$ for $t \ge 1$ and $F_{ij}(0)=1-S_{ij}(0)$ for $t =0$.  In the Laplace domain, we establish the relation ${{\tilde F}_{ij}}\left( s \right) =1+ \left( {{e^{ - s}} - 1} \right){{\tilde S}_{ij}}\left( s \right)$. Substituting the relation into Eq.\ref{eq19}, we have
\begin{equation}\label{eq20}
\tilde S_{ij}\left( s \right) = \frac{{1 - \tilde F_{ij}\left( s \right)}}{{1 - {e^{ - s}}}} = \frac{{\tilde P_{jj}\left( s \right) - \tilde P_{ij}\left( s \right) + {\delta _{ij}}}}{{\left( {1 - {e^{ - s}}} \right)\tilde P_{jj}\left( s \right)}}.
\end{equation}
The MFPT is calculated as
\begin{align}\label{eq21}
\left\langle {{T_{ij}}} \right\rangle & = \mathop {\lim }\limits_{s \to 0} {\tilde S_{ij}}\left( s \right) = \mathop {\lim }\limits_{s \to 0} \frac{{{{\tilde P}_{jj}}\left( s \right) - {{\tilde P}_{ij}}\left( s \right) + {\delta _{ij}}}}{{\left( {1 - {e^{ - s}}} \right){{\tilde P}_{jj}}\left( s \right)}} \nonumber \\ &= \frac{1}{{{P_j}\left( \infty  \right)}}\mathop {\lim }\limits_{s \to 0} \left[ {{{\tilde P}_{jj}}\left( s \right) - {{\tilde P}_{ij}}\left( s \right) + {\delta _{ij}}} \right].
\end{align}
In the last step in Eq.(\ref{eq21}), we have used the result of Eq.(\ref{eq10}). Utilizing Eq.(\ref{eq9}), we calculate the limit in Eq.(\ref{eq21}) by using the L'H\^opital rule, which leads to (see Appendix \ref{appen3} for details) 
\begin{equation}\label{eq22}
\left\langle T_{ij}\right\rangle  = 
\left\{ 
\begin{array}{lr}
\frac{1}{P_{j}\left(\infty\right)}\left(Z_{jj}-Z_{ij} \right) + \sum_{k=1}^{n} \left(Z_{ik} - Z_{jk}\right), & i \ne j, \\
\frac{1}{{P_j}\left(\infty\right)}, & i = j.
\end{array} 
\right.
\end{equation}

It is also useful to quantify the ability of a process to explore
the whole network \cite{PhysRevE.80.065104,PhysRevLett.109.088701}. For this purpose, we define $T(j)$ as the global MFPT to the target node $j$, averaging over the starting node $i$,
\begin{eqnarray}\label{eq23}
T(j) = \frac{1}{n} \sum\limits_{i = 1}^n {\left\langle T_{ij}\right\rangle}.
\end{eqnarray}

\section{Results on various networks}
\subsection{Ring networks}

\begin{figure}
	\centerline{\includegraphics*[width=1.0\columnwidth]{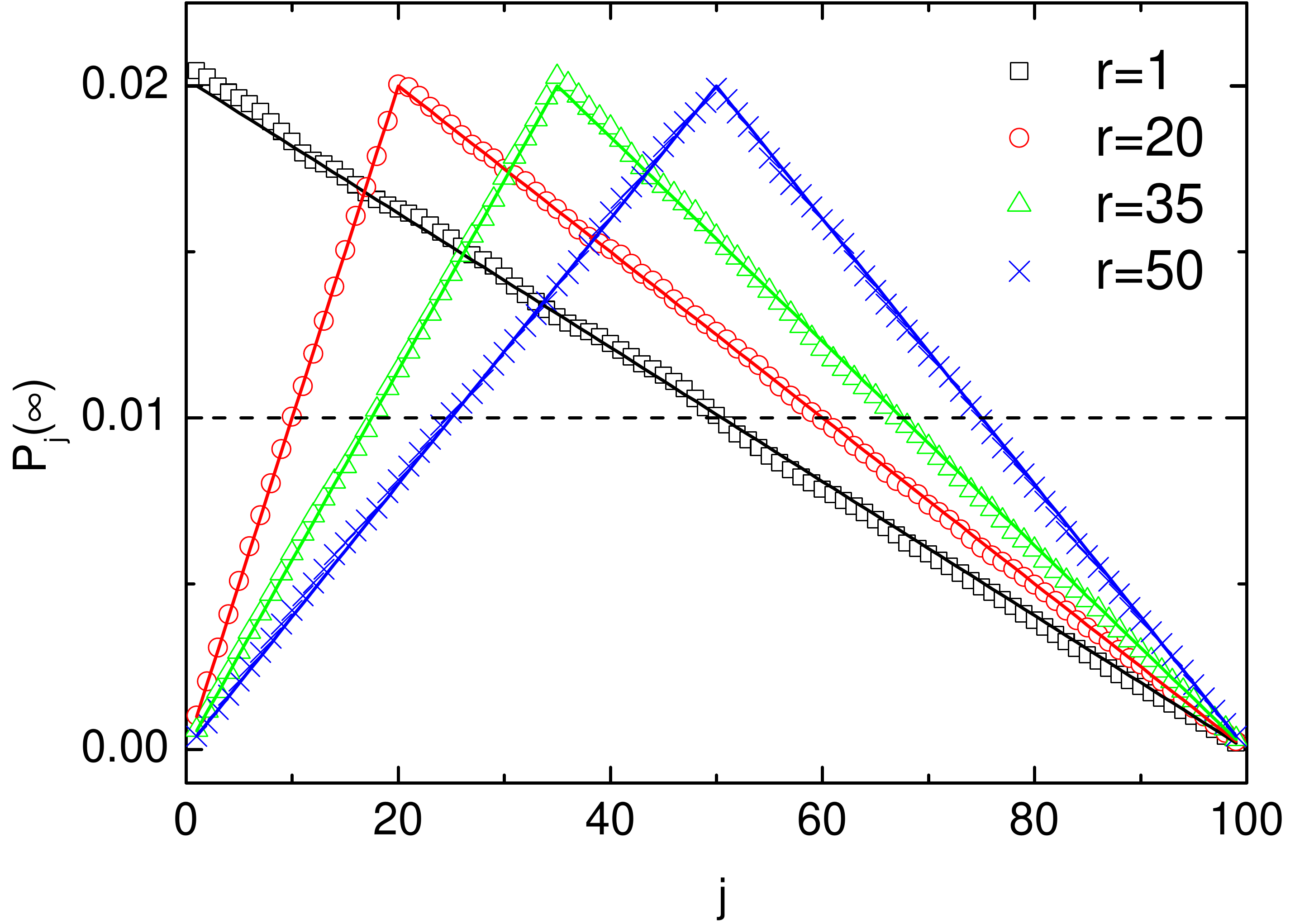}}
	\caption{The stationary occupation probability on a ring of size $N=100$ with different resetting node $r$. The only observable node is set to be the last node. Symbols and lines represent the simulation and theoretical results, respectively. Dashed line indicates the result of standard random walks for comparison. }\label{fig2}
\end{figure}

\begin{figure*}
	\centerline{\includegraphics*[width=1.6\columnwidth]{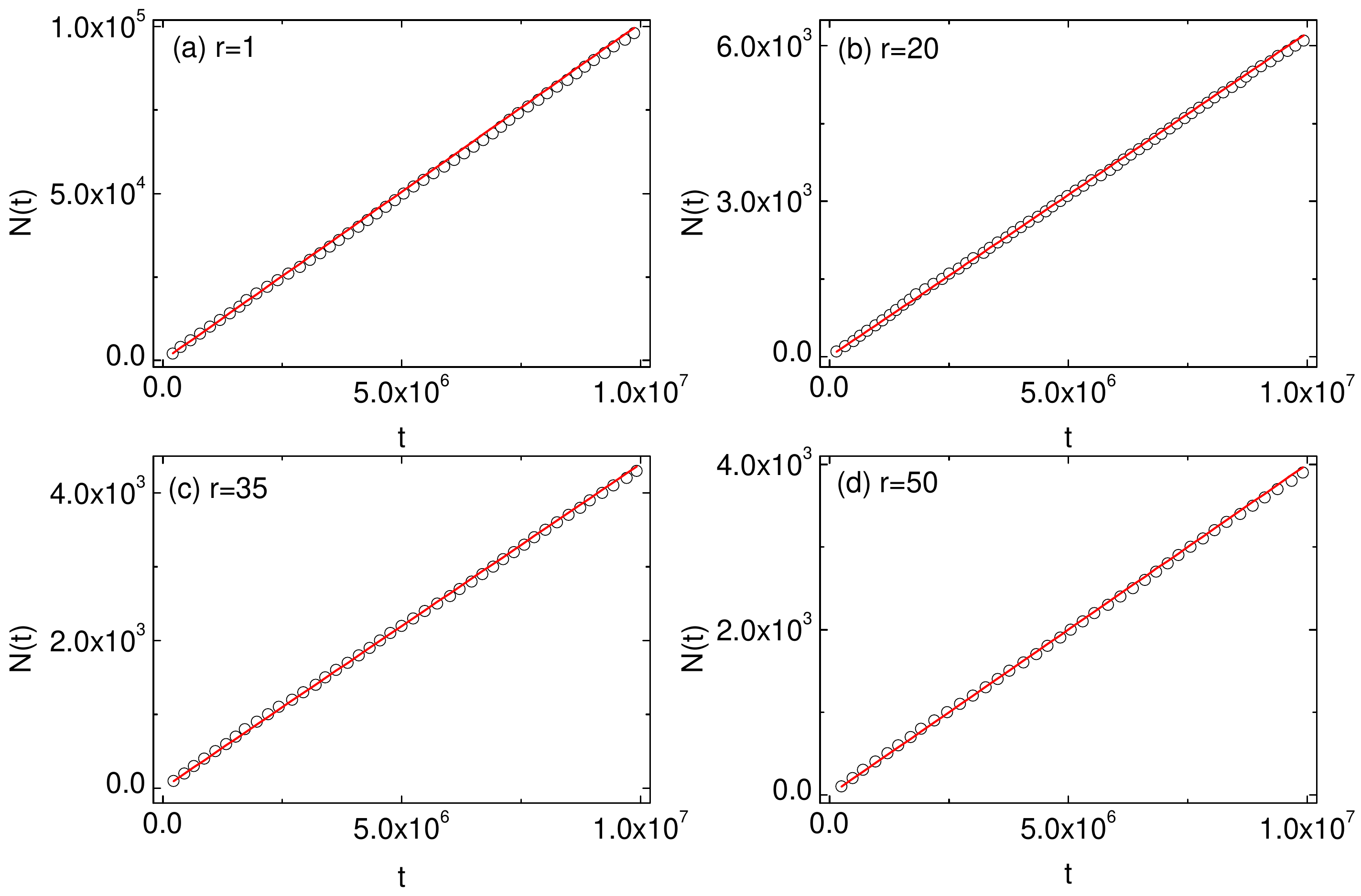}}
	\caption{The number of resets as a function of time on a ring of size $N=100$ with different resetting node $r$. Symbols and lines represent the simulation and theoretical results, respectively. From (a)-(d), the resetting node is set as, $r=1,20,35,50$, respectively.}\label{fig3}
\end{figure*}

We consider a ring network with size $N$, and choose one of nodes as the only observable node. Without loss of generality, we set the last node as the only observable node. The transition matrix $\textbf{Q}$ between non-observable nodes can be written as 
\begin{equation}\label{eq24}
\textbf{Q} = \left( {\begin{array}{*{20}{c}}
	0&{1/2}&{}&{}&{}\\
	{1/2}&0&{1/2}&{}&{}\\
	{}&{1/2}& \ddots & \ddots &{}\\
	{}&{}& \ddots & \ddots &{1/2}\\
	{}&{}&{}&{1/2}&0
	\end{array}} \right).
\end{equation}
We can see that $\textbf{Q}$ is a symmetric tridiagonal matrix. $\textbf{I}_n-\textbf{Q}$ is also a symmetric tridiagonal matrix, and its inverse $\textbf{Z}$ can be obtained explicitly \cite{usmani1994inversion},
\begin{equation}\label{eq25}
{Z_{ij}} = \frac{{2\min \left\{ {i,j} \right\}\left( {N - \max \left\{ {i,j} \right\}} \right)}}{N}.
\end{equation}

According to Eq.(\ref{eq11}), the stationary occupation probability is given by
\begin{equation}\label{eq26}
{P_j}\left( \infty  \right) = 
\left\{ 
\begin{array}{lr}
\frac{2j}{rN},& j \le r,\\
\frac{2\left({N-j}\right)}{N\left( {N - r} \right)}, & j \ge r,
\end{array} 
\right.
\end{equation}
from which we can see that ${P_j}\left( \infty  \right) $ firstly increases linearly for $j \in [1, r]$ and then decreases linearly for $j \in [r, N-1]$. In Fig.\ref{fig2}, we show $P_j(\infty)$ as a function of $j$ on a ring of size $N=100$, where the resetting node $r$ is set to be four different nodes, $r=1, 20, 35, 50$, respectively. The theoretical and simulation results are represented by the lines and symbols, respectively, and they are in well agreement. In all simulations, we have used $10^7$ time steps to estimate the stationary occupation probability.

According to Eq.(\ref{eq17}), the growth rate of $\mathcal{N}(t)$ is given by 
\begin{equation}\label{eq27}
\sigma = \frac{1}{(N-r)r},
\end{equation}
from which we can see that the growth rate is minimized when $r=N/2$. The result can be intuitively interpreted as follows. When $r=N/2$ (noting that the observable node is the $N$th node), the geodesic distance between the resetting node and the observable node is maximal, so that the MFPT between them is maximized (see Eq.\ref{eq33}) and thus the growth rate of $\mathcal{N}(t)$ is minimized. In Fig.\ref{fig3}, we compare the simulation result with the theory for different resetting node: $r=1, 20, 35, 50$, as shown by the symbols and lines, respectively. Our theory completely predicts the linear behavior of $\mathcal{N}(t) \sim t $.

According to Eq.(\ref{eq22}), the MFPT is given by
\begin{widetext}
	\begin{equation}\label{eq28}
	\left\langle {{T_{ij}}} \right\rangle  = \left\{ \begin{array}{lr}
	\frac{2}{{N{P_j}\left( \infty  \right)}}\left[ {j\left( {N - j} \right) - i\left( {N - j} \right)} \right] + i\left( {N - i} \right) - j\left( {N - j} \right),&i < j,\\
	\frac{2}{{N{P_j}\left( \infty  \right)}}\left[ {j\left( {N - j} \right) - j\left( {N - i} \right)} \right] + i\left( {N - i} \right) - j\left( {N - j} \right),&i > j,\\
	\frac{1}{{{P_j}\left( \infty  \right)}},&i = j.
	\end{array} \right.
	\end{equation}
\end{widetext}
and the global MFPT is given by
\begin{widetext}
\begin{equation}\label{eq29}
T\left( j \right) = \left\{ \begin{array}{lr}
\frac{r}{{2Nj}}\left[ {N + \left( {N - 2} \right)\left( {N - j} \right)j} \right] + \frac{1}{{6N}}\left( {N - 1} \right)\left[ {{N^2} + \left( {1 - 6j} \right)N + 6{j^2}} \right],& j \le r,\\
\frac{{\left( {N - r} \right)}}{{2N\left( {N - j} \right)}}\left[ {N + \left( {N - 2} \right)\left( {N - j} \right)j} \right] + \frac{1}{{6N}}\left( {N - 1} \right)\left[ {{N^2} + \left( {1 - 6j} \right)N + 6{j^2}} \right],&j \ge r.
\end{array} \right.
\end{equation}
\end{widetext}

\begin{figure*}
	\centerline{\includegraphics*[width=1.8\columnwidth]{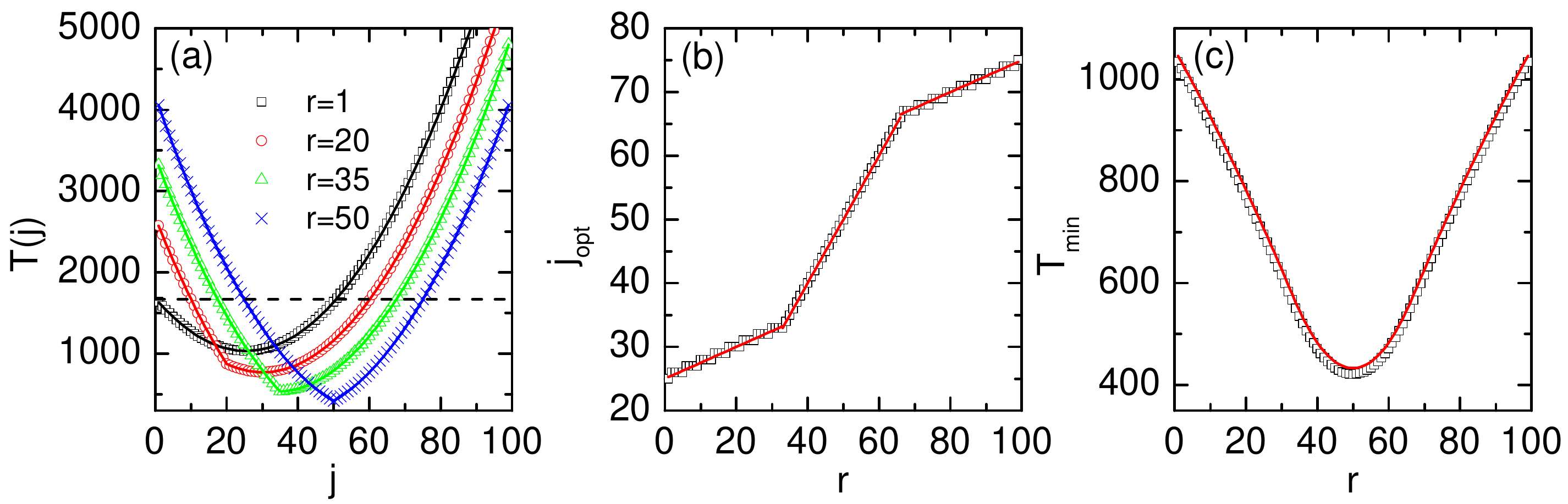}}
	\caption{Results on a ring of size $N=100$. (a) The global MFPT as a function of target node $j$ with different resetting node $r$. Dashed line indicates the result of SRW for comparison. (b) The optimal target node $j_{\rm {opt}}$ as a function of resetting node $r$. (c) The minimal global MFPT, $T_{\min}$, as a function of resetting node $r$. Symbols and solid lines represent the simulation and theoretical results, respectively. }\label{fig4}
\end{figure*}

In Fig.\ref{fig4}a, we show the global MFPT as a function of the target node $j$ for different resetting node: $r=1,20,35,50$. Symbols and lines represent the simulation and theoretical results, respectively. In all simulations, we have used $10^4$ averages to estimate the MFPT between arbitrary two nodes. We can see that $T(j)$ exhibits a non-monotonic change with $j$. There exists an optimal $j=j_{\rm {opt}}$ for which $T(j)$ is minimal, $T_{\min}=T(j_{\rm {opt}})$. We will derive the expressions of $j_{\rm {opt}}$ and $T_{\min}$, and show they are dependent on the resetting node and the size of the ring.

In addition, for the standard random walks (SRW) on a ring of size $N$, the MFPT from node $i$ to node $j$ is given by \cite{montroll1965random}
\begin{equation}\label{eq33}
{\left\langle {{T_{ij}}} \right\rangle _{\rm{SRW}}} = \left\{ \begin{array}{lr}
{d_{ij}}\left( {N - {d_{ij}}} \right),&i \ne j,\\
N,&i = j,
\end{array} \right.
\end{equation}
where $d_{ij}=\min \left\{|i-j|, N-|i-j|  \right\}$ is the geodesic distance between node $i$ and node $j$.  
From Eq.(\ref{eq33}), we obtain the global MFPT 
\begin{equation}\label{eq34}
T{\left( j \right)_{\rm{SRW}}} = \frac{1}{6}\left( {{N^2} + 5} \right). 
\end{equation}  
For comparison, we show the global MFPT for the SRW (that is independent of the target node $j$) in Fig.\ref{fig4}a, indicated by the dashed line. We observe a wide range of $j$ values for which $T(j)$ is smaller than that in the case of SRW.  
 
In the limit of $N \to \infty$, Eq.(\ref{eq29}) can be approximately written as
\begin{equation}\label{eq30}
T\left( j \right) = \left\{ \begin{array}{l}
\frac{{r\left( {N - j} \right)}}{2} + \frac{1}{6}\left[ {{N^2} + \left( {1 - 6j} \right)N + 6{j^2}} \right],j \le r,\\
\frac{{\left( {N - r} \right)j}}{2} + \frac{1}{6}\left[ {{N^2} + \left( {1 - 6j} \right)N + 6{j^2}} \right],j \ge r.
\end{array} \right.
\end{equation}
As shown in Fig.\ref{fig4}(a), $T(j)$ exhibits a non-monotonic variation with $j$. At an optimal value of $j=j_{\rm {opt}}$, $T(j)$ is minimized, $T_{\min}=T(j_{\rm {opt}})$. From Eq.\ref{eq30}, we find that both $j_{\rm {opt}}$ and $T_{\min }$ are dependent on the resetting node $r$ and the size of ring $N$, given by
\begin{equation}\label{eq31}
{j_{\rm {opt}}} = \left\{ \begin{array}{lr}
\frac{{N + r}}{4},&r \le \frac{N}{3},\\
r,& \frac{N}{3} \le r \le \frac{{2N}}{3},\\
\frac{{2N + r}}{4}, &r \ge \frac{{2N}}{3},
\end{array} \right.
\end{equation}
and 
\begin{equation}\label{eq32}
{T_{\min }} = \left\{ \begin{array}{lr}
\frac{1}{{48}}\left[ {5{N^2} + N\left( {8 - 6r} \right) - 3{r^2}} \right], &r \le \frac{N}{3},\\
\frac{1}{6}\left[ {{N^2} + N\left( {1 - 3r} \right) + 3{r^2}} \right],&\frac{N}{3} \le r \le \frac{{2N}}{3},\\
\frac{1}{{48}}\left[ { - 4{N^2} + 4N\left( {2 + 3r} \right) - 3{r^2}} \right], & r \ge \frac{{2N}}{3}.
\end{array} \right.
\end{equation}
In Fig.\ref{fig4}b and Fig.\ref{fig4}c, we show the values of $j_{\rm {opt}}$ and $T_{\min}$ as a function of the resetting node $r$, respectively. The theoretical and simulation results are represented by the lines and symbols, respectively, and they are in well agreement.   $j_{\rm {opt}}$ is a piecewise linear function of $r$ and increases with $r$. $T_{\min}$ is a piecewise quadratic function of $r$, and shows a minimum at $r=N/2$. 
We should note that the theory and simulations depart a little bit more in the optimal region, as shown Fig.\ref{fig4}c. The error is essentially originated from the discarding of the subleading orders of $N$ in Eq.(\ref{eq30} approximated by Eq.(\ref{eq29})). Furthermore, $T_{\min}$ is minimized in the optimal region, rendering the error is more obviously shown. We have tested other sizes of rings as well, and found that the relative error between the theory and simulations decreases with increasing $N$.

\subsection{2d square lattices}

\begin{figure}
	\centerline{\includegraphics*[width=0.6\columnwidth]{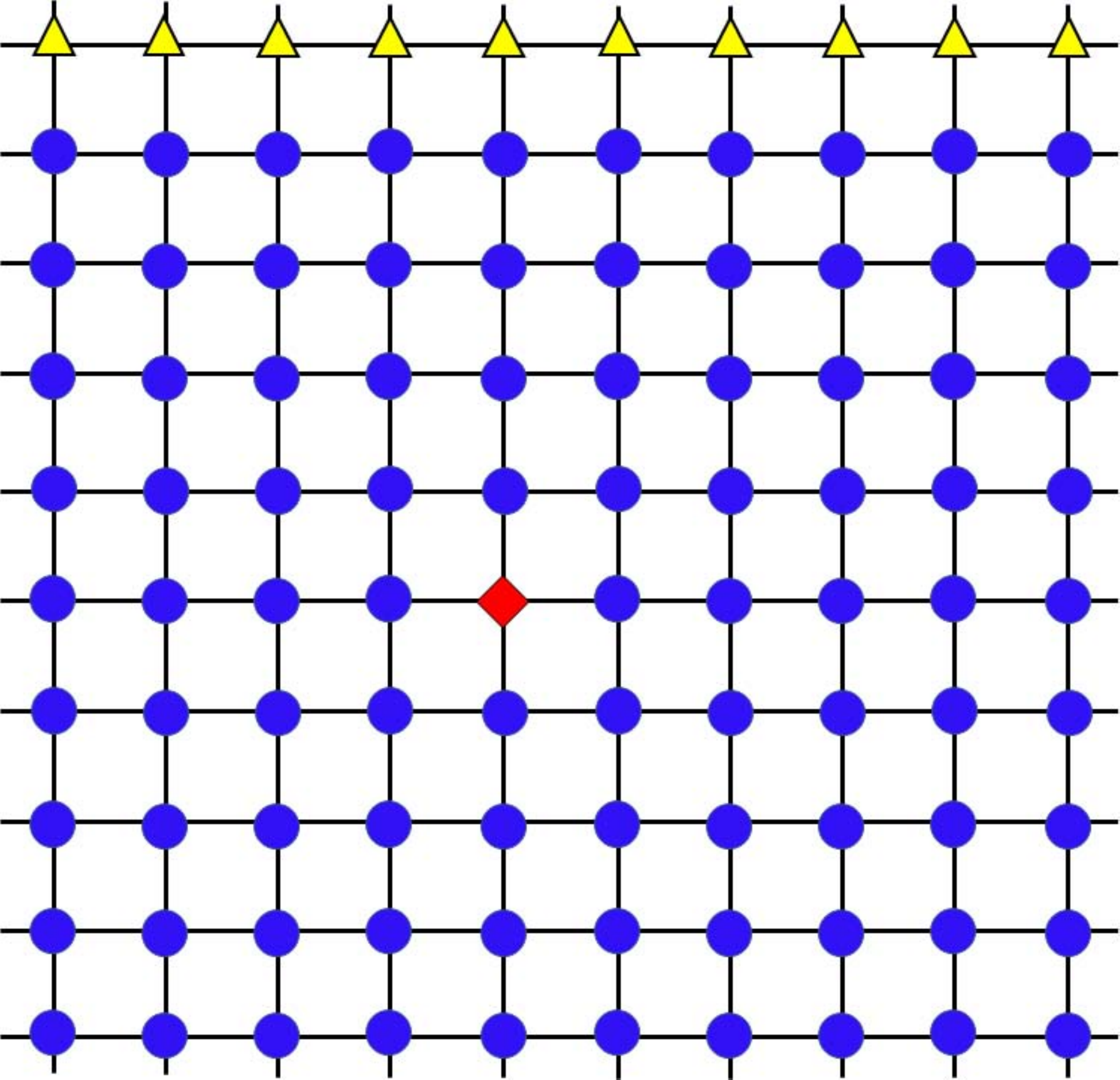}}
	\caption{A two-dimensional square lattice with size $N=10 \times 10$, in which we set a  resetting node ($\diamond$), and 10 observable nodes ($\triangle$). Nodes are numbered from left to right and from bottom to top.}\label{fig5}
\end{figure}

\begin{figure}
	\centerline{ \includegraphics*[width=1.0\columnwidth]{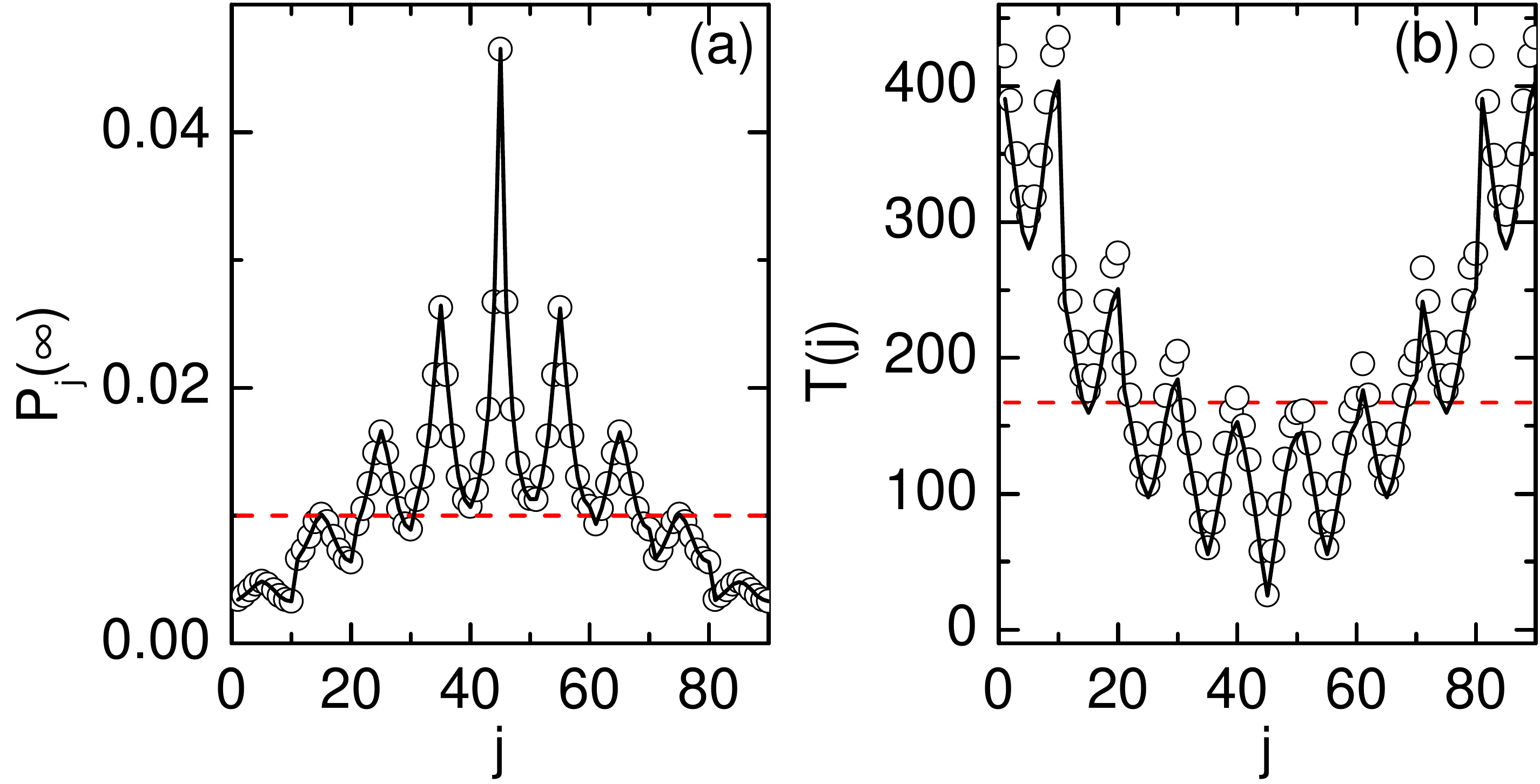}}
	\caption{Results on a two-dimensional square lattice shown in Fig.\ref{fig5}. (a) The stationary occupation probability as a function of the label of node $j$. (b) The global MFPT as a function of the target node $j$. Solid lines and symbols represent the theoretical and simulation results, respectively. Dashed lines indicate the results of the SRW. }\label{fig6}
\end{figure}

\begin{figure}
	\centerline{ \includegraphics*[width=1.0\columnwidth]{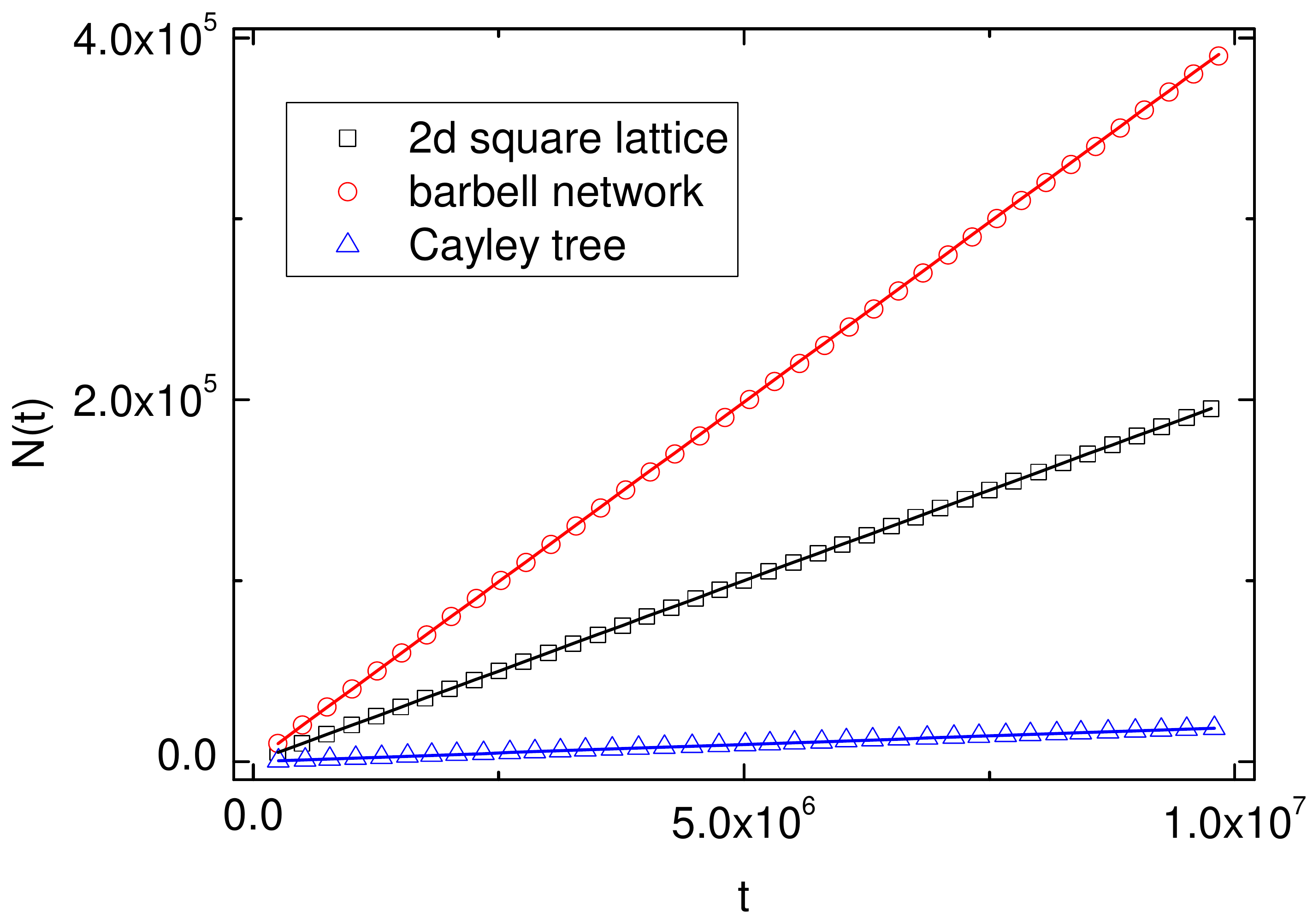}}
	\caption{The number of resets as a function of time on a 2d square lattice (see Fig.\ref{fig5}), a barbell network (see Fig.\ref{fig7}), and a finite-size Cayley tree (see Fig.\ref{fig9}). Symbols and lines represent the simulation and theoretical results, respectively.}\label{fign}
\end{figure}

We consider a two-dimensional square lattice of size $N=10\times 10$ with periodic boundaries in both dimensions, as shown in Fig.\ref{fig5}. We first choose one side on the lattice as observable sites, i.e., 10 nodes as observable nodes ($\triangle$). We then choose a node as the only resetting node ($\diamond$). The results on the 2d lattice is shown in Fig.\ref{fig6}. Similar to the results on 1d rings, both the stationary occupation probability and the global MFPT become non-homogeneous due to the resetting processes, in contrast to the SRW on regular graphs. The stationary occupation probabilities of the nodes close to the resetting node are much higher, and the global MFPT of these nodes are much smaller. In addition, we find that the growth rate of $\mathcal{N}(t)$ with $t$ is about $\sigma \approx  0.02$. The simulation can also reproduce the growth rate very well, as shown in Fig.\ref{fign}.

\subsection{Barbell networks}

\begin{figure}
	\centerline{\includegraphics*[width=0.8\columnwidth]{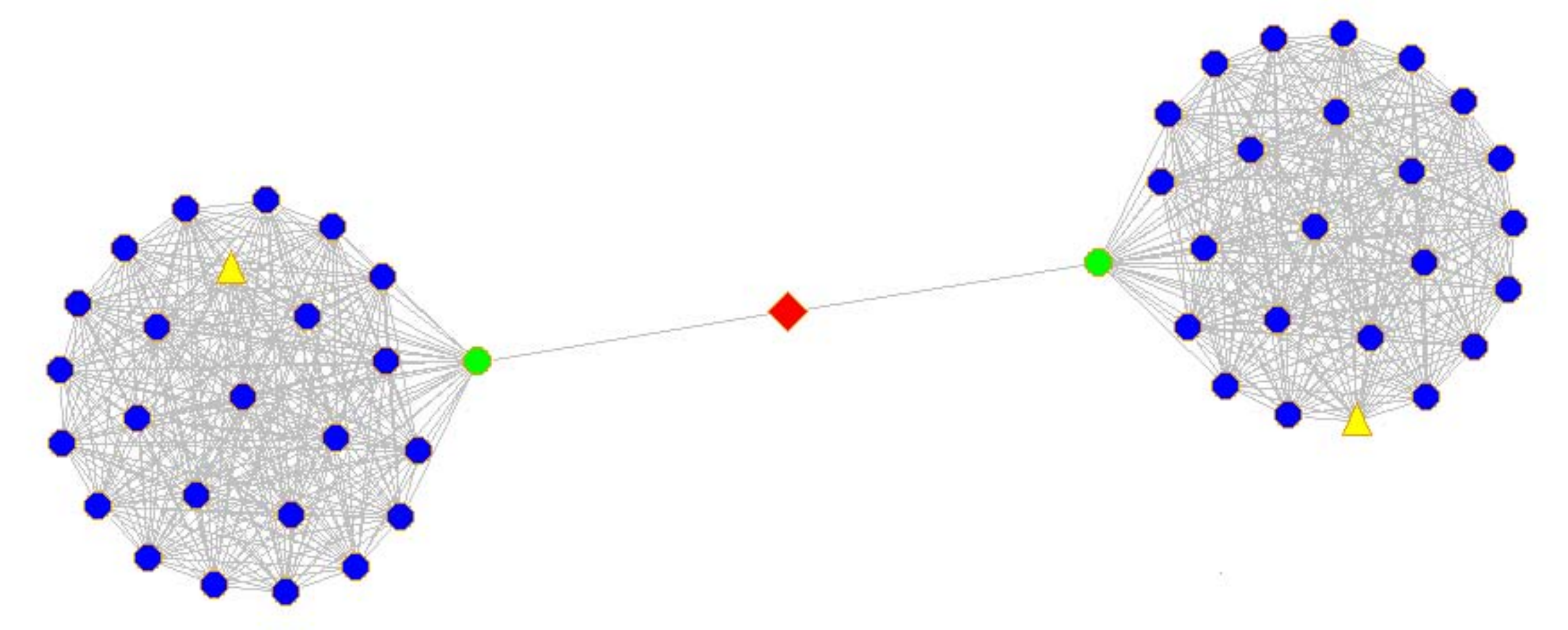}}
	\caption{A barbell network with size $N=51$ composed of two fully connected
		subgraphs (of 25 nodes each) connected by a bridge node. We set the bridge node as  the resetting node ($\diamond$), and choose one node in each subgraph as the observable node ($\triangle$), but different from two nodes connected to the bridge node. }\label{fig7}
\end{figure}

\begin{figure*}
	\centerline{\includegraphics*[width=1.6\columnwidth]{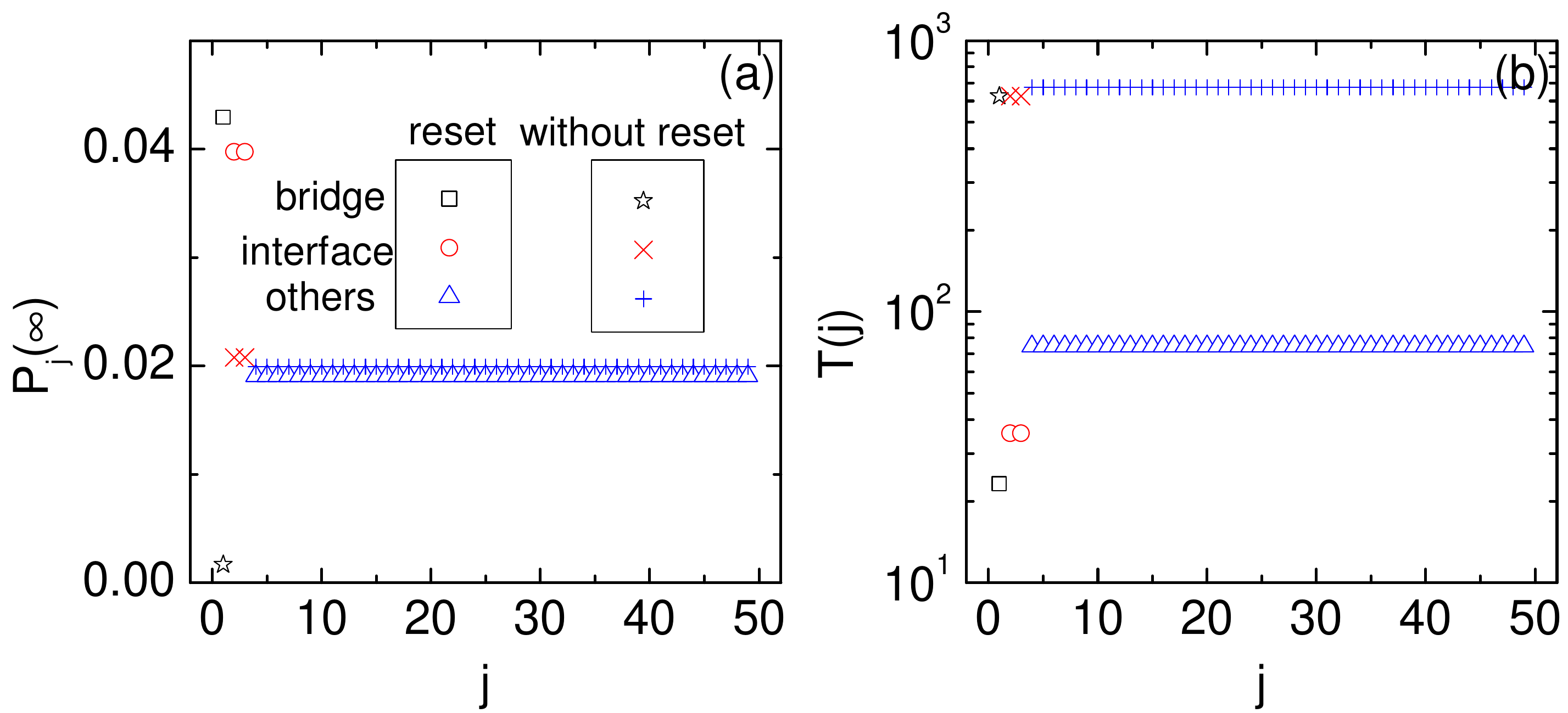}}
	\caption{Results on a barbell network with shown in Fig.\ref{fig7}. (a) The stationary occupation probability as a function of node $j$. (b) The global MFPT as a function of the target node $j$. For comparison, we also show the results of no resetting, corresponding to the case of SRW. Three types of nodes, including a bridge node, two interface nodes and other nodes, are represented by different symbols. All results are obtained from the theory.  }\label{fig8}
\end{figure*}

We consider a barbell network with size $N=51$ that is composed of two fully connected
subgraphs (of 25 nodes each) connected by a bridge node, as shown in Fig.\ref{fig7}. We set the bridge node as the resetting node ($\diamond$), and randomly choose one node in each subgraph as the observable node ($\triangle$), but different from two nodes connected to the bridge node (named as the interface nodes). For the SRW, the particle is prone to be trapped in either of subgraphs, and thus make the global search difficult \cite{rosvall2008maps,jeub2015think,masuda2009impact}. This difficulty can be overcome by resetting the particle to the bridge node. It is thus expected that the presence of the first-passage resetting promotes the efficiency of global search. In Fig.\ref{fig8}, we summarize the results in the barbell network obtained from our theory. 
We have performed the corresponding simulations for verifying the theory as well. We find that the maximum relative error between the theory and simulations is less than $0.9 \%$ and $2.3 \%$ for $P_j(\infty)$ and $T_j$, respectively.
For a clear visualization, the simulation results do not shown in Fig.\ref{fig8}.  However, for comparison, we also show the results of SRW in Fig.\ref{fig8}. In the presence of the first-passage resetting, the stationary occupation probability of the resetting node is the largest, that is slightly larger than that of the two interface nodes, but is about twice larger than those of other non-observable nodes. This is contrast to the case in SRW, in which the stationary occupation probability of the resetting node is the smallest. The global MFPT for each node in the case of the first-passage resetting is much less than that in the case of the SRW. In detail, for the bridge node, the interface nodes, and the other non-observable nodes, the global MFPT in SRW is about 27, 17, and 9 times larger than that in random walks with the first-passage resetting, respectively. Therefore, the first-passage resetting is advantageous to the search in such networks. Our theory also shows the number of resets is a linearly increasing function of time, giving the growth rate, $\sigma \approx 0.04$, that is verified by simulations (see Fig.\ref{fign}).

\subsection{Cayley trees}
\begin{figure}
	\centerline{\includegraphics*[width=0.8\columnwidth]{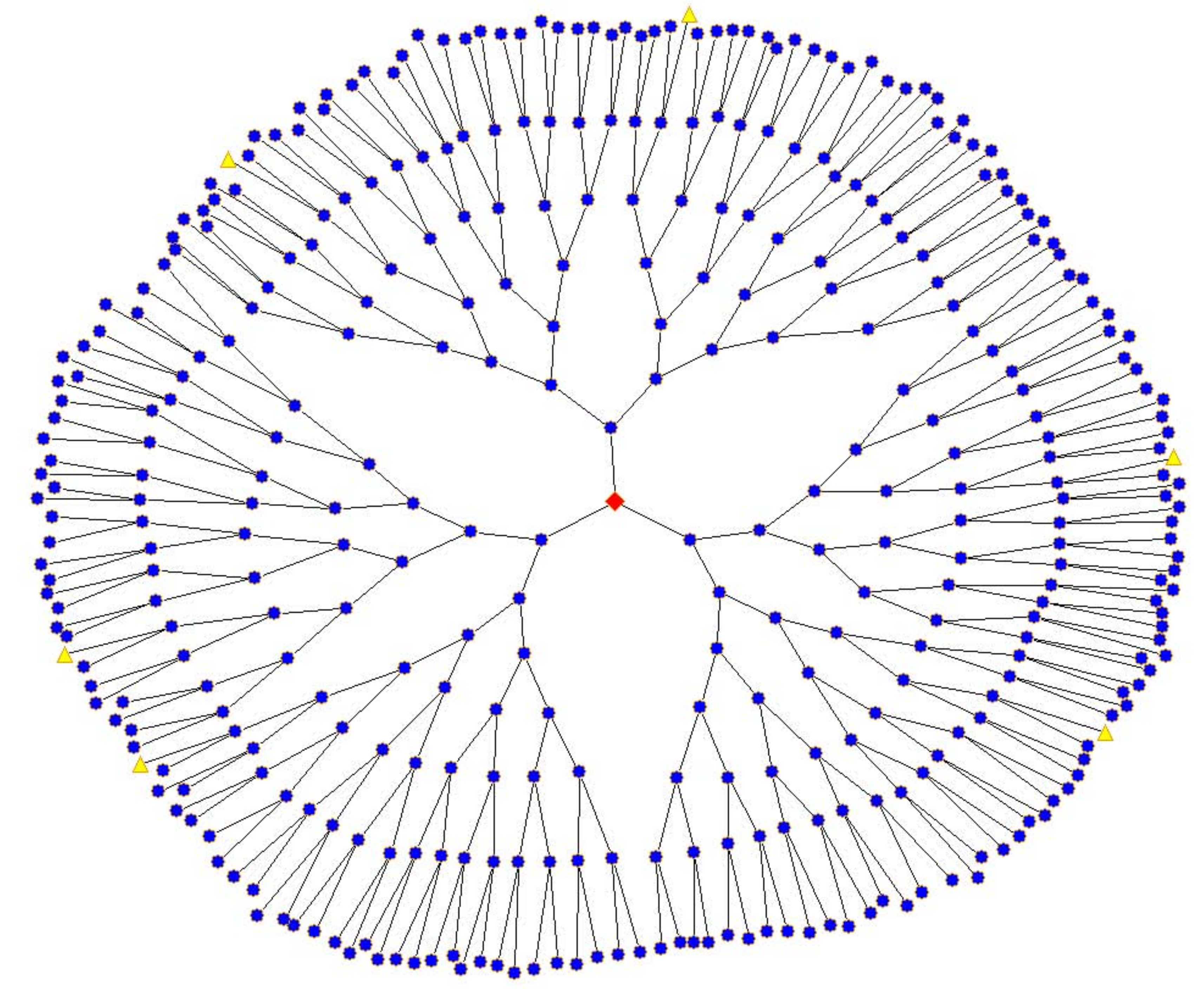}}
	\caption{A Cayley tree with coordination number $z = 3$ and 7 shells ($N = 382$ nodes). The root node is set to be the resetting node ($\diamond$), and 6 leaf nodes to be observable nodes ($\triangle$). Nodes are numbered by the order of shells.}\label{fig9}
\end{figure}

\begin{figure*}
	\centerline{\includegraphics*[width=1.6\columnwidth]{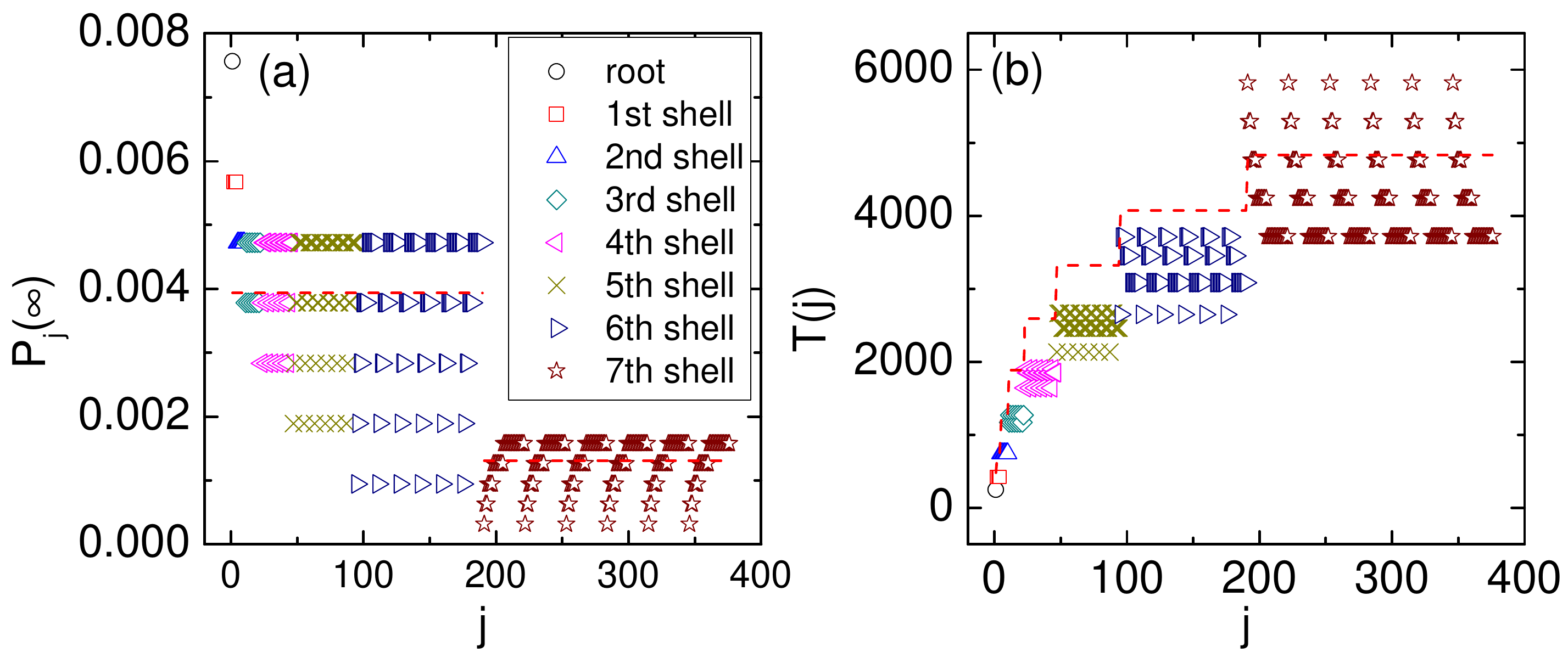}}
	\caption{Results on the Cayley tree shown in Fig.\ref{fig9}. (a) The stationary occupation probability as a function of node $j$. (b) The global MFPT as a function of the target node $j$. For comparison, we also show the results of SRW (dashed lines). All results are obtained from the theory. }\label{fig10}
\end{figure*}

We now consider a finite Cayley tree of coordination number
$z=3$ and composed of $n=7$ shells (see Fig.\ref{fig9}). The nodes in
the last shell have degree 1, whereas the other nodes have
degree $z$. The root node is set to be the resetting node ($\diamond$), and 6 leaf nodes to be observable nodes ($\triangle$). 
In Fig.\ref{fig10}, we show the results on the Cayley tree. Due to the resetting to the root node, on the one hand, the stationary occupation probabilities of the nodes in inner shells are relatively higher. On the other hand, the stationary occupation probabilities become more diverse compared to the SRW (see dashed line in Fig.\ref{fig10}).  On average, the stationary occupation probabilities of the nodes whose shell is less than 5 is higher to those in the SRW. The global MFPT of most of nodes are smaller than those in the standard random walks, except for a small amount of nodes in the outermost shell. We should note that all the results in Fig.\ref{fig10} are obtained from the theory. We also made simulations and find the maximum relative error between the theory and simulations is less than $3.6 \%$ and $2.1 \%$ for $P_j(\infty)$ and $T_j$, respectively. In addition, our theory predicts that the growth rate of $\mathcal{N}(t)$ with $t$ is $\sigma \approx  1.89 \times 10^{-3}$, in agreement with the simulations (see Fig.\ref{fign} for comparison).

\section{Conclusions}
We have explored the effect of the first-passage resetting on the random walks on general networks. In our model, a set of nodes are firstly chosen as the observable nodes, and a node as the only resetting node. The walker is reset instantaneously to the resetting node whenever it reaches either of observable nodes, and the random walk process is restarted. Based on the renewal theory, we have derived the exact expressions of the stationary occupation probability, the average number of resets, and the mean first-passage time between arbitrary two nodes. Interestingly, we find that all the quantities can be expressed by the so-called fundamental matrix $\textbf{Z}$, that is the inverse of an identity matrix minus the transition matrix between non-observable nodes. We have demonstrated our results on various networks, including ring networks, two-dimensional square lattices, barbell networks, and finite Cayley trees. The results showed that the first-passage resetting brings the model to a nonequilibrium steady state, and can accelerate the global MFPT with respect to the standard random walks. For ring networks and two-dimensional square lattices, the stationary occupation probabilities for each node become nonhomogeneous and the global MFPT can be reduced for these nodes close to the resetting node. For barbell networks, it is striking that the global MFPT for all nodes can be reduced, embodying the advantage of the first-passage resetting on such networks with community structures \cite{fortunato2010community}. For finite Cayley trees, the global MFPT can be reduced for most of nodes, except for a small amount of nodes in the outermost shell. These results may open up a novel way to exploring complex networks. In the future, it is interesting to generalize our model to the cases of multiple resetting nodes and multiplex networks \cite{bianconi2018multilayer}.

\begin{acknowledgments}
This work is supported by the National Natural Science Foundation of China (Grants No. 11875069, No 61973001) and the Key Scientific Research Fund of Anhui Provincial Education Department under (Grant No. KJ2019A0781).	
\end{acknowledgments}

\appendix
\section{Derivation of $P_j$($\infty$)}\label{appen1}
Subsituting Eq.\ref{eq9} to Eq.\ref{eq10}, we have
\begin{align}\label{eqa1}
{P_j}\left( \infty  \right) &= \mathop {\lim }\limits_{s \to 0} \left( {1 - {e^{ - s}}} \right)\left[ {{{\tilde G}_{ij}}\left( s \right) + \frac{{{{\tilde F}_{i\textbf{o}}}\left( s \right)}}{{1 - {{\tilde F}_{r\textbf{o}}}\left( s \right)}}{{\tilde G}_{rj}}\left( s \right)} \right] \nonumber \\&=\mathop {\lim }\limits_{s \to 0} \left( {1 - {e^{ - s}}} \right)\frac{{{{\tilde F}_{i\textbf{o}}}\left( s \right)}}{{1 - {{\tilde F}_{r\textbf{o}}}\left( s \right)}}{{\tilde G}_{rj}}\left( s \right)
\end{align}
Since ${{\tilde F}_{i\textbf{o}}}\left( 0 \right) = \sum_{t = 0}^\infty  {{F_{i\textbf{o}}}} \left( t \right) = 1$ (implying that the hitting probability from any non-observable node to observable nodes is always one), the limit in Eq.\ref{eqa1} has the form of $0/0$, and thus we then apply the L'H\^opital rule to calculate the limit, which leads to 
\begin{align}\label{eqa2}
{P_j}\left( \infty  \right) =  - \frac{{{{\tilde F}_{i\textbf{o}}}\left( 0 \right){{\tilde G}_{rj}}\left( 0 \right)}}{{{{\tilde F'}_{r\textbf{o}}}\left( 0 \right)}} =  - \frac{{{{\tilde G}_{rj}}\left( 0 \right)}}{{{{\tilde F'}_{r\textbf{o}}}\left( 0 \right)}},
\end{align}
where ${{\tilde G}_{rj}}\left( 0 \right)$ is given by Eq.\ref{eq6} and Eq.\ref{eq12},
\begin{align}\label{eqa3}
{{\tilde G}_{rj}}\left( 0 \right) = {\left[ {{{\left( {{\textbf{I}_n} - \textbf{Q}} \right)}^{ - 1}}} \right]_{rj}} = {Z_{rj}},
\end{align}
and ${{{\tilde F'}_{r\textbf{o}}}\left( 0 \right)}$ is the derivative of ${{{\tilde F}_{r\textbf{o}}}\left( s \right)}$ with respect to $s$ at $s=0$, given by Eq.\ref{eq7}
\begin{align}\label{eqa4}
{\tilde{F}'_{r\textbf{o}}}\left( 0 \right)=- \sum\limits_{k = 1}^m {{{\left[ {{{\left( {{\textbf{I}_n} - \textbf{Q}} \right)}^{ - 2}}\textbf{R}} \right]}_{rk}}}. 
\end{align}
Since ${\left( {{\textbf{I}_n} - \textbf{Q}} \right)^{ - 1}} = {\textbf{I}_n} + \textbf{Q} + {\textbf{Q}^2} +  \cdots $, we have 
\begin{align}\label{eqa5}
	\sum\limits_{k = 1}^m {{{\left[ {{{\left( {{\textbf{I}_n} - \textbf{Q}} \right)}^{ - 1}}\textbf{R}} \right]}_{ik}}}  = \sum\limits_{k = 1}^m {{{ {\left( {\textbf{R} + \textbf{Q}\textbf{R} + {\textbf{Q}^2}\textbf{R} +  \cdots } \right)} }_{ik}}} .
\end{align}
Subsituting Eq.\ref{eq4} into Eq.\ref{eqa5}, we have
\begin{align}\label{eqa6}
	\sum\limits_{k = 1}^m {{{\left[ {{{\left( {{\textbf{I}_n} - \textbf{Q}} \right)}^{ - 1}}\textbf{R}} \right]}_{ik}}}  = \sum\limits_{k = 1}^m {\sum\limits_{t = 0}^\infty  {{F_{i{o_k}}}\left( t \right)  } }= \sum\limits_{t = 0}^\infty  {{F_{i\textbf{o}}}\left( t \right)}  = 1.
\end{align}
Furthermore, 
\begin{widetext}
	\begin{align}\label{eqa8}
		\sum\limits_{k = 1}^m {{{\left[ {{{\left( {{\textbf{I}_n} - \textbf{Q}} \right)}^{ - 2}}\textbf{R}} \right]}_{rk}}}  &= \sum\limits_{l = 1}^n {\sum\limits_{k = 1}^m {{{\left[ {{{\left( {{\textbf{I}_n} - \textbf{Q}} \right)}^{ - 1}}} \right]}_{rl}}{{\left[ {{{\left( {{\textbf{I}_n} - \textbf{Q}} \right)}^{ - 1}}\textbf{R}} \right]}_{lk}}} } \nonumber \\ &= {\sum\limits_{l = 1}^n {{{\left[ {{{\left( {{\textbf{I}_n} - \textbf{Q}} \right)}^{ - 1}}} \right]}_{rl}}} },		
	\end{align}
\end{widetext}
where we have utilized the result of Eq.\ref{eqa6} in the last line. 
Substituting Eq.\ref{eqa8} into Eq.\ref{eqa4}, combining the definition of Eq.\ref{eq12}, we have 
\begin{align}\label{eqa9}
{\tilde{F}'_{r\textbf{o}}}\left( 0 \right)=-\sum\limits_{k = 1}^m {{{\left[ {{{\left( {{\textbf{I}_n} - \textbf{Q}} \right)}^{ - 1}}} \right]}_{rk}} }=  - \sum\limits_{k = 1}^m {{Z_{rk}}} .
\end{align}
Subsituting Eq.\ref{eqa3} and Eq.\ref{eqa9} into Eq.\ref{eqa2}, we obtain the result of Eq.\ref{eq11}.

\section{Asymptotic behavior of $\mathcal{N}(t)$ in the long time}\label{appen2}
The asymptotic behavior of $\mathcal{N}(t)$ in the long time can be obtained by calculating $\mathcal{\tilde {N}}(s)$ in the limit of $s \to 0$. To the end, we consider the limit, 
\begin{align}\label{eqb1}
\mathop {\lim }\limits_{s \to 0} {\left( {1 - {e^{ - s}}} \right)^2}\mathcal{\tilde N}\left( s \right).
\end{align}
According to Eq.\ref{eq15} and Eq.\ref{eq7}, Eq.\ref{eqb1} becomes
\begin{align}\label{eqb2}
	\mathop {\lim }\limits_{s \to 0} {\left( {1 - {e^{ - s}}} \right)^2}\mathcal{\tilde N}\left( s \right)=\mathop {\lim }\limits_{s \to 0} \frac{{1 - {e^{ - s}}}}{{1 - {{\tilde F}_{r\textbf{o}}}\left( s \right)}}.
\end{align}
Since Eq.\ref{eqb2} has the form of $0/0$, we apply the L'H\^opital rule to calculate the limit, which leads to 
\begin{align}\label{eqb3}
	\mathop {\lim }\limits_{s \to 0} {\left( {1 - {e^{ - s}}} \right)^2}\mathcal{\tilde N}\left( s \right)=- \frac{1}{{{{\tilde F'}_{r\textbf{o}}}\left( s \right)}}.
\end{align}
Utilizing the result of Eq.\ref{eqa9}, we obtain the asymptotic behavior of $\mathcal{\tilde N}\left( s \right) $ in the limit of $s \to 0$, 
\begin{align}\label{eqb4}
	\mathcal{\tilde N}\left( s \right)=\sigma {\left( {1 - {e^{ - s}}} \right)^{ - 2}}
\end{align}
where $\sigma  =1/{\sum\nolimits_{k = 1}^m {{{{Z}}_{rk}}} }$ is the reciprocal of the sum of the $r$th row of the fundamental matrix. Performing the inverse transform for Eq.\ref{eqb4}, we immediately obtain the result of Eq.\ref{eq16}.

\section{Derivation of the MFPT}\label{appen3}
In order to derive the MFPT from Eq.\ref{eq21}, we need to calculate the limit  
\begin{align}\label{eqc1}
\mathop {\lim }\limits_{s \to 0} \left[ {{{\tilde P}_{jj}}\left( s \right) - {{\tilde P}_{ij}}\left( s \right)} \right]
\end{align}
Subsituting Eq.\ref{eq9} into Eq.\ref{eqc1}, we have 
\begin{widetext}
\begin{align}\label{eqc2}
\mathop {\lim }\limits_{s \to 0} \left[ {{{\tilde P}_{jj}}\left( s \right) - {{\tilde P}_{ij}}\left( s \right)} \right] = \mathop {\lim }\limits_{s \to 0} \left\{ {{{\tilde G}_{jj}}\left( s \right) - {{\tilde G}_{ij}}\left( s \right) + \frac{{{{\tilde G}_{rj}}\left( s \right)}}{{1 - {{\tilde F}_{r\textbf{o}}}\left( s \right)}}\left[ {{{\tilde F}_{j\textbf{o}}}\left( s \right) - {{\tilde F}_{i\textbf{o}}}\left( s \right)} \right]} \right\}.
\end{align}
\end{widetext}
According to Eq.\ref{eq6} and Eq.\ref{eq12}, we have
\begin{align}\label{eqc3}
\mathop {\lim }\limits_{s \to 0} \left[ {{{\tilde G}_{jj}}\left( s \right) - {{\tilde G}_{ij}}\left( s \right)} \right] = {Z_{jj}} - {Z_{ij}}.
\end{align}
Since the limit 
\begin{align}\label{eqc4}
\mathop {\lim }\limits_{s \to 0} \frac{{{{\tilde G}_{rj}}\left( s \right)}}{{1 - {{\tilde F}_{r\textbf{o}}}\left( s \right)}}\left[ {{{\tilde F}_{j\textbf{o}}}\left( s \right) - {{\tilde F}_{i\textbf{o}}}\left( s \right)} \right]
\end{align}
has the form of $0/0$, we apply the L'H\^opital rule to calculate the limit, which yields to 
\begin{widetext}
\begin{align}\label{eqc5}
\mathop {\lim }\limits_{s \to 0} \frac{{{{\tilde G}_{rj}}\left( s \right)}}{{1 - {{\tilde F}_{r\textbf{o}}}\left( s \right)}}\left[ {{{\tilde F}_{j\textbf{o}}}\left( s \right) - {{\tilde F}_{i\textbf{o}}}\left( s \right)} \right] =- \frac{{{{\tilde G}_{rj}}\left( 0 \right)}}{{  {{\tilde F'}_{r\textbf{o}}}\left( 0 \right)}}\left[ {{{\tilde F'}_{j\textbf{o}}}\left( 0 \right) - {{\tilde F'}_{i\textbf{o}}}\left( s \right)} \right]
\end{align}
\end{widetext}
Substituting Eq.\ref{eqa9} into Eq.\ref{eqc5}, we obtain
\begin{widetext}
\begin{align}\label{eqc6}
\mathop {\lim }\limits_{s \to 0} \frac{{{{\tilde G}_{rj}}\left( s \right)}}{{1 - {{\tilde F}_{r\textbf{o}}}\left( s \right)}}\left[ {{{\tilde F}_{j\textbf{o}}}\left( s \right) - {{\tilde F}_{i\textbf{o}}}\left( s \right)} \right ] & = \frac{{{Z_{rj}}}}{{\sum\nolimits_{k = 1}^n {{Z_{rk}}} }}\sum\nolimits_{k = 1}^n {\left( {{Z_{ik}} - {Z_{jk}}} \right)} \nonumber \\ &= {P_j}\left( \infty  \right)\sum\nolimits_{k = 1}^n {\left( {{Z_{ik}} - {Z_{jk}}} \right)}  
\end{align}
\end{widetext}
where we have used Eq.\ref{eq11} in the last line of Eq.\ref{eqc6}. Substituting Eq.\ref{eqc3} and Eq.\ref{eqc6} into Eq.\ref{eqc2}, we obtain
\begin{widetext}
\begin{align}\label{eqc7}
\mathop {\lim }\limits_{s \to 0} \left[ {{{\tilde P}_{jj}}\left( s \right) - {{\tilde P}_{ij}}\left( s \right)} \right] = {Z_{jj}} - {Z_{ij}} + {P_j}\left( \infty  \right)\sum\nolimits_{k = 1}^n {\left( {{Z_{ik}} - {Z_{jk}}} \right)} 
\end{align}
\end{widetext}
Substituting Eq.\ref{eqc7} into Eq.\ref{eq21}, we immediately obtain the MFPT given by Eq.\ref{eq22}.


\end{document}